\documentclass[pra,letterpaper,twocolumn]{revtex4-1}   
\usepackage{amsbsy,amssymb,amsmath,bm,bbold} 
\usepackage{graphicx,color,epsfig,rotate} 
\usepackage{fancyhdr} 
\usepackage{epstopdf}
\usepackage{csquotes}
\usepackage{float}
\usepackage[colorlinks=true,linkcolor=blue,citecolor=blue,urlcolor=blue]{hyperref}

\begin{document}
	

	
	\title{Photonic Crystal Based Ultra-Sensitive Interferometric Sensor with Spatial Resolution  upto 1 nm}
	
	\author{Snehashis Sadhukhan,$^1$ Arnab Laha,$^{1,2}$ Abhijit Biswas,$^1$ and Somnath Ghosh$^{2,}$}
	\email{somiit@rediffmail.com}
	\affiliation{\vspace{0.3cm}$^1$Institute of Radiophysics and Electronics, University of Calcutta, Kolkata-700009, India\\
		$^2$Department of Physics, Indian Institute of Technology Jodhpur, Rajasthan-342037, India}

	\begin{abstract}
	We report a very high precision interferometric sensor with resolution up to $\sim\lambda/1024$, exploiting hollow photonic bandgap waveguide-based geometry for the first time. Here sensing has been measured by a complete switching in the direction of the outgoing beam, owing to transverse momentum oscillation phenomena. Using a ${1.32\,\mu}$m source and core-width of ${7.25\,\mu}$m, a complete switching cycle is obtained even due to a small change of $\sim$1 nm in the core-width. Using hollow-core photonic bandgap waveguide, Talbot effect, revivals of the initial phase, oscillation in the transverse momentum along with multi-mode interference served as the backbone of the design. The ultra-sensitive multi-mode interferometric sensor based on photonic crystals will certainly open up a paradigm shift in interferometer based sensing technologies toward device-level applications in photonic sensing/switching and related precision measurement systems. 
	\end{abstract}
	
	\maketitle
	
	
	Interferometry is a technique where different characteristics of light waves and the matters they interact with are measured precisely and examined intensively, primarily to sense displacement, vibration, distance, etc. In the optical domain, interferometric techniques have a huge number of applications including sensing \cite{sensorappl01,sensorappl02}, modulation, \cite{modappl01,modappl02} and switching technologies \cite{switchappl01,switchappl02}. Lately, the demand for precise measurement of many physical quantities to test optical systems a wide range of optical interferometers is built. In the year 2001, Y. B. Ovchinnikov et al. put forward their new concept of such interferometers using parallel plane waveguides \cite{ovchinnikov01}. Talbot effect \cite{berry96}, revivals of initial phase \cite{ovchinnikov00,bryngdahl73}, oscillation in the transverse momentum \cite{ovchinnikov01} along with multi mode interference (MMI) \cite{bryngdahl73,ulrich75} served as the back-bone of their design, implementing which they achieved $\sim\lambda/9$ fringe spacing. In 2008, R. M. Jenkins et al. \cite{jenkins08} made a significant improvement of this technique, where exploiting the same methodology with further optimization in the operating condition they were able to reach $\sim\lambda/130$ fringe spacing. Such a fine order of measurement was indeed useful in applications in the study of flow rate not limited to aerodynamics to visualize the flow in wind tunnels \cite{book1,chevalerias57}, heat transfer to measure pressure, density temperature changes in gases \cite{book3}. Moreover, similar technique can be further useful as  an efficient device to be used in electro-optic modulator that not limited to fiber-optic communication, but can be implemented in monolithic circuits that provide a very high-bandwidth electro-optic amplitude and phase signals in a region of operation of over few gigahertz frequencies and as holographic interferometric device \cite{holography} even in macro/micro interferometry.

	To fit the demand, they came up with a design that contains two parallel fully reflecting metal mirrors together acting as a planar waveguide and light wave entering into it at an angle $\alpha$, which in turn excites different modes inside the waveguide depending on the angle $\alpha$ \cite{jenkins08}. On meeting the condition for multi mode propagation and interference inside the waveguide for the input wave, revivals of the initial phase along with the oscillation in the transverse momentum and self-imaging phenomena are observed. As a consequence of these multiple phenomena, the output beam has been governed to move to and fro between the angles  $+\alpha$ to  $-\alpha$ whenever the guide width is consequently changed. In their work, they considered two fully reflecting gold coated mirrors, one fixed and other mounted on an actuator to form the waveguide, and then they coupled a $1.32\,\mu m$ Nd:YAG laser source using a single-mode, polarization maintaining optical fiber to the planar waveguide. They demonstrated how a source of $1.32\,\mu m$ at an incident angle $0.14\,rad$ excites two different transverse electric(TE) modes (e.g., TE$_1$ and TE$_2$ modes) inside the interferometric structure of width $8\,\mu m$ and length $50\,mm$, and produces the spatial resolution of $10\,nm$. Such a specially configured interferometeric structure was novel and better suitable for application in precision optical measurements or switching when compared with other of its type such as Fabry-P\'erot or Michelson interferometer, since it produces much smaller fringe spacing than $\lambda/2$.
	
	Of late, in many instruments such as optical splicers, modulators, switches, etc. higher resolution needs to be realized alongside increased accuracy for far and wide applications. But with a view to achieve higher resolution using the existing setup is restricted  due to the presence of large attenuation encountered by higher order modes emerged from the reflection in between two metal mirrors.
	
\begin{figure}[t]
		\includegraphics[width=8.5cm]{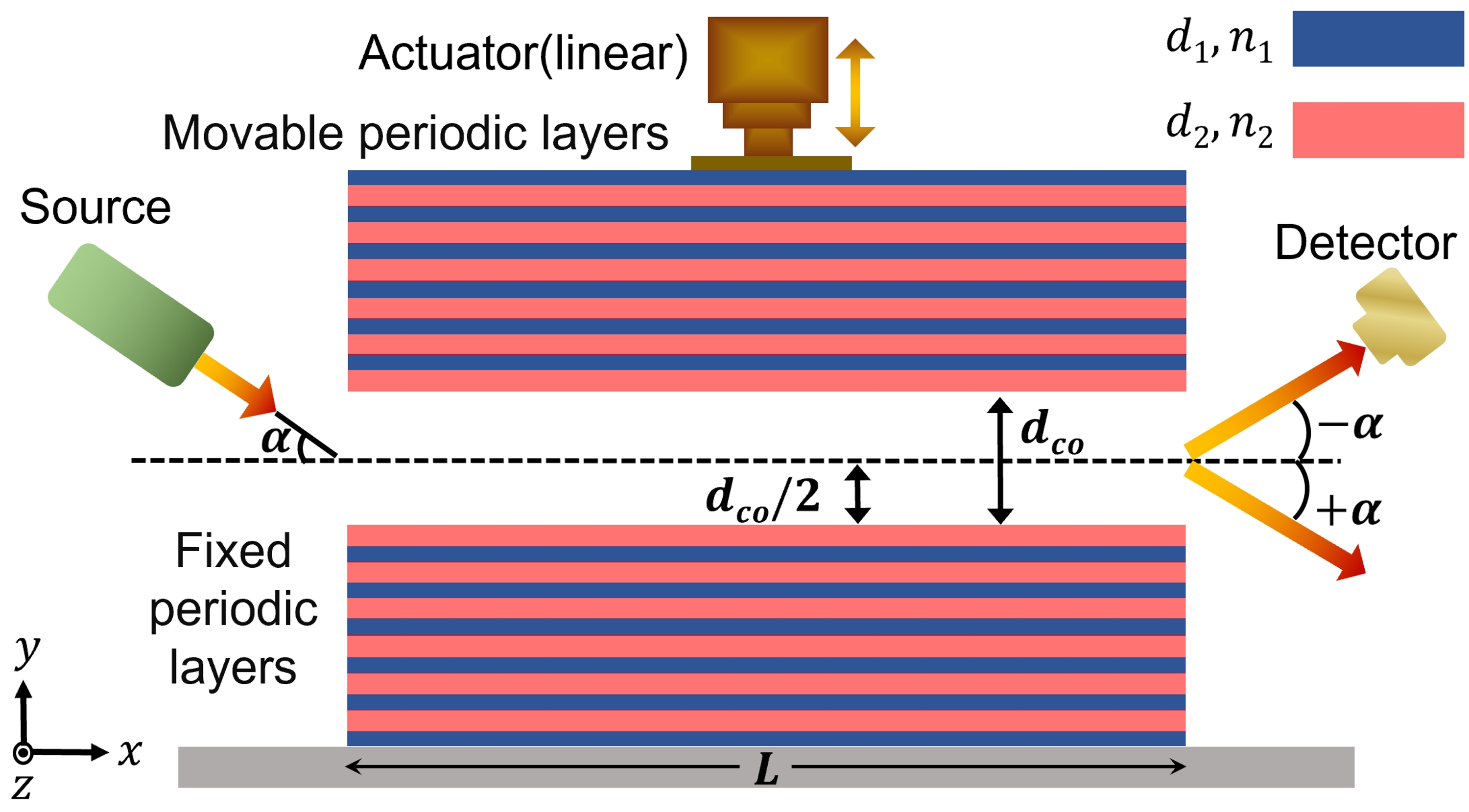}
		\caption{Schematic diagram of photonic crystal based interferometric sensor with the associated 1D photonic crystals, actuator, input laser and detector configuration. $\alpha$ represents the input angle, $d_{co}$ is the core width and {$d_1$,$n_1$ and $d_2$,$n_2$} are the thicknesses and refractive indices of low and high index materials respectively. $L$ is the length of the photonic crystal.}
		\label{fig1}
	\end{figure}

	In this letter, we propose a new prototype of photonic crystal based interferometric sensor to achieve the resolution significantly enhanced beyond the limitations of former designs.
	Exploring the light guiding mechanism, the interplay between multimode guidance and loss minimization of higher order modes has been judiciously implemented to achieve the targeted resolution.
	Focusing on the interaction between lower order mode with a higher order mode, and keeping in mind that the relative loss should be minimum along with a maximum difference in the order of selected modes, the prototype has been proposed. To sense a small displacement, as small as $1\,nm$, we design our interferometric device using 1D photonic crystals. In our design, we guide the wave utilizing the bandgap supported by 1D periodic structure, that consists of $6$ bi-layers in both the arms (as shown in Fig. \ref{fig1}), where the spatial widths of the low index $(n_1=3.2)$ and high index $(n_2=3.8)$ materials  have been chosen to be $0.095\,\mu m$ and $1\,\mu m$ respectively. The core width has been optimised at $7.25\,\mu m$. One arm is fixed and the other is attached to a high precision actuator that is used to change the core width. A Nd: YAG source has been taken into account that emits a wavelength of ${1.32\,\mu m}$ and assumed to be coupled with the waveguide using a single mode fiber with mode diameter $5\,\mu m$ \cite{henderson76}.
	
\begin{figure}[b!]
		\includegraphics[width=8.5cm]{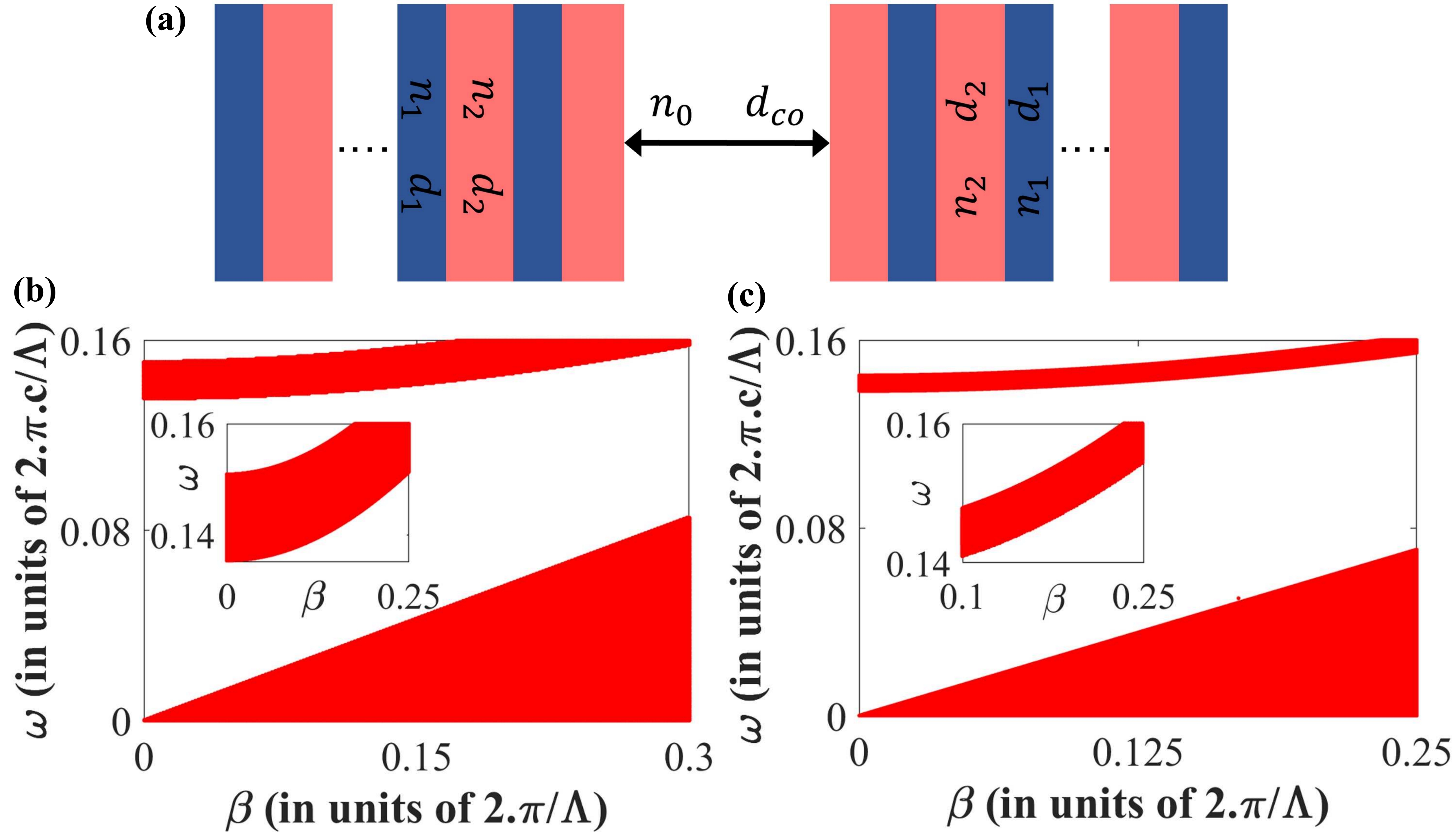}
		\caption{(a) Basic schematic of photonic crystal based bandgap guided structure with alternating layers. Band diagram (TE polarisation) for periodic layers optimised in the proposed design defined by (b) $n_1=3.2,\,n_2=3.8,\,d_1=1.00\,\mu m,\,d_2=0.95\,\mu m$, and (c) $n_1=3.4,\,n_2=3.66,\,d_1=1.00\,\mu m,\,d_2=0.95\,\mu m$ respectively. $\Lambda$ represents the periodicity of the medium ( $\Lambda=d_1+d_2$ ).}
		\label{fig2}
	\end{figure}

	On account of guiding a light of wavelength 1.32 $\mu m$ light using Photonic Crystals, bandgap engineering has been performed to achieve the corresponding stop band by modulating the spatial  periodicity ($\Lambda$) and refractive indices$(n)$ of the periodic cladding. Fig. \ref{fig2}(a) shows the basic waveguide structure to realize our proposed interferometric sensor. We optimize two sets of operating parameters that control the bandgap of the 1D photonic crystals for which the corresponding bandgap plots have been shown in Figs. \ref{fig2}(b) and (c) (the corresponding choices of parameters have been indicated in the caption of Fig. \ref{fig2}) \cite{yeh76,book2}. To achieve few bandgap guided lowest order modes with minimum loss, we have optimized the bandgap structure with parameters as $n_1=3.2,\,n_2=3.8,\,d_1=1.00\,\mu m,\,d_2=0.95\,\mu m$ shown in Fig. \ref{fig2}(a).
	The number of bi-layers ($N$), in the spatially periodic cladding, governs the modal loss inside the waveguide. In our proposed structure, we choose $N=6$, which results in a confinement loss \cite{pal07} of 0.0005 dB/m for (TE${_0}$) mode and loss value even lower than 
	0.01 dB/m for the supported higher order modes.
	\begin{figure}[t]
		\includegraphics[width=8.5cm]{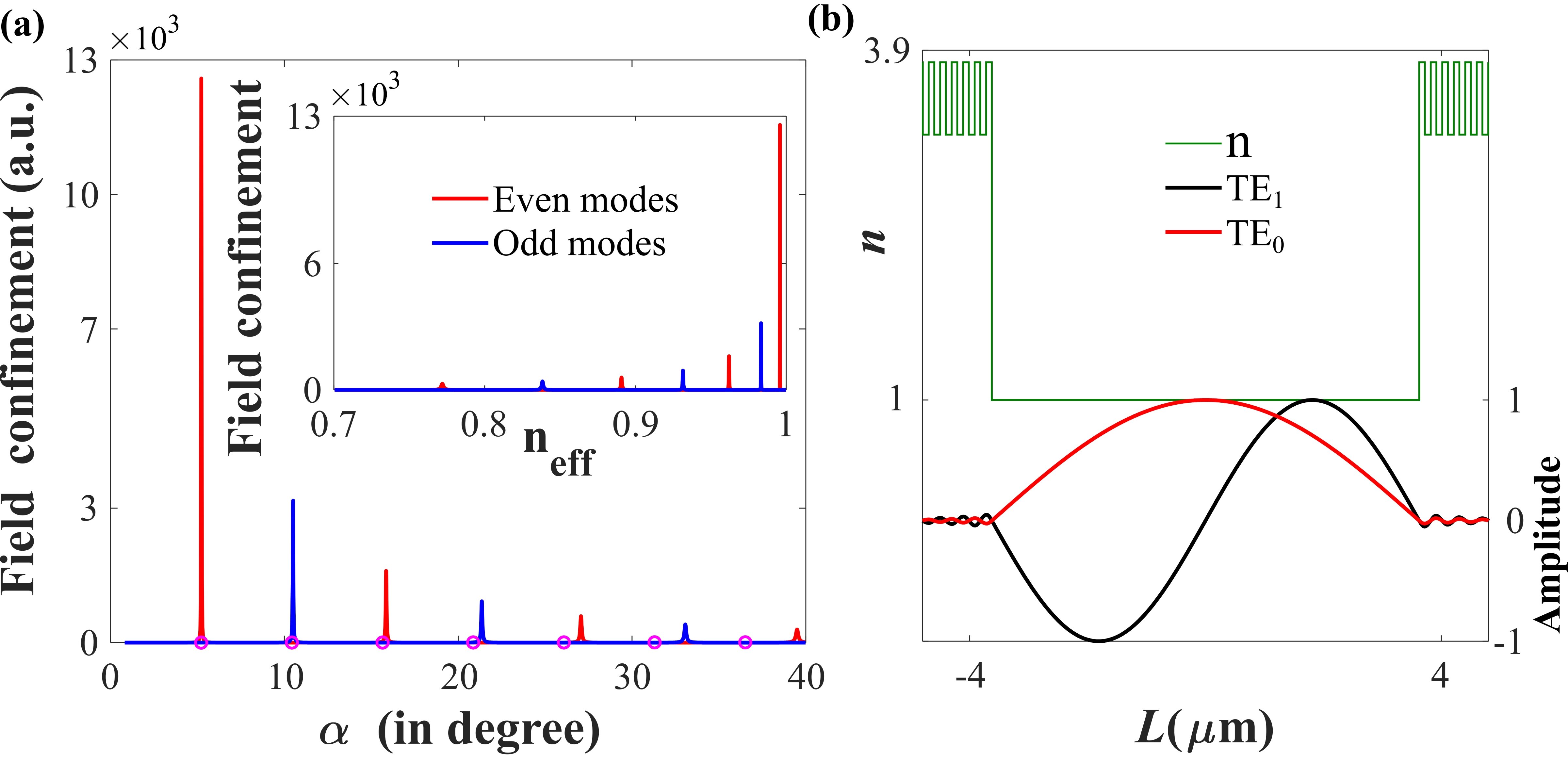}
		\caption{(a) Field confinements of the supported modes with corresponding input angles ($\alpha$). Inset shows the variation with effective mode index ($n_{eff}$). (b) Optimised refractive index (${n}$) profile of the waveguide and the profiles of the supported lowest order (TE$_0$,TE$_1$) modes.}
		\label{fig3}
	\end{figure}
	Further, optimizing the core width ($d_{co}$) is an important requirement for individual field confinement and relative loss among the supported modes which dictates the intensity contrast of modes inside the guide. Accordingly, $d_{co}$ has been chosen so that the fundamental mode (TE$_0$) produces maximum field confinement inside the guide, when fibre and waveguide axes are perfectly aligned (i.e., $\alpha=0$), which turns out to be $7.25\,\mu m$.
	Suitably, the single mode polarization maintaining fiber, corresponding to waveguide core width of $7.25\,\mu m$ is found to have a mode field diameter of $5\,\mu m$ for maximum power coupling from fiber to the waveguide \cite{henderson76}.
	
	In Fig. \ref{fig3}(a), we have shown the field confinements of the supported modes by the waveguide for an increasing value of input angles ($\alpha$). In the inset we have shown the similar field confinements with effective index of the respective modes. In Fig. \ref{fig3}(a), the red and blue curves indicate the confinements of the even and odd modes of different order, respectively. The corresponding input angles ($\alpha$) to excite the individual supported modes have been marked by violet circular markers. The mode profiles of the two lowest order even (red line) and odd (black line) modes supported by the optimized waveguide geometry have also been displayed in Fig. \ref{fig3}(b) along with the optimized index profile ($n$)   of the waveguide ( shown by green curve ). 
	The incident angle $\alpha$ can be approximated to produce $m^{th}$ mode inside the guide as 
	\begin{equation}
	\alpha= (m+1)\frac{\lambda}{2d_{co}}\times\frac{180}{\pi}.
	\label{angle}
	\end{equation}
	Here $\lambda$ is the operating wavelength ($=1.32\,\mu m$) of the input from Nd:YAG source.
	Thus, for an incident angle $\alpha$ for $m=0$ only TE$_0$ will get excite. We can dissolve the phase constant $k$ of any incoming wave into two components $k_x= k\cos(\alpha)$ and $k_y= k\sin(\alpha)$. 
	Now, on exciting two modes (say, $u$ and $v$, with $\{u,v\}\in m$) inside the waveguide with proper choice of $\alpha$, the phase difference between them, after propagation of distance $L$ along the axial (i.e. $x$) direction, can written as 
	\begin{equation}
	\Phi_{uv}(L)=L(k_{xu}-k_{xv})=\frac{\pi\lambda L\left\{(v+1)^2-(u+1)^2\right\}}{4d_{co}^2},
	\label{psipq}
	\end{equation}
	where, $k_{xu}$ and $k_{xv}$  represent $x$ component of the phase coefficient of mode $u$ and $v$ respectively. Here, the change in phase due to small change in core width can be obtained from Eq. \ref{psipq} as
	\begin{equation}
	\delta\Phi= -\frac{(v+1)^2-(u+1)^2}{2}\left(\frac{L\pi\lambda}{d_{co}^3}\right)\delta d_{co}.
	\label{dphi}
	\end{equation}
	A complete switching of output deflected beam is observed when $\delta \Phi$ accumulates the value of $2\pi$. Thus, considering $\delta\Phi=2\pi$ Eq. \ref{dphi} can be rewritten as
	\begin{equation}
	\delta d_{co}= -\frac{4}{(v+1)^2-(u+1)^2}\left(\frac{d_{co}^3}{L\lambda}\right).
	\label{dco}
	\end{equation}
	Eq. \ref{dco} signifies that the amount of displacement experienced by the upper arm of the waveguide that will make the output beam to swing back to its initial direction.
	\begin{figure}[b!]
		\includegraphics[width=8.5cm]{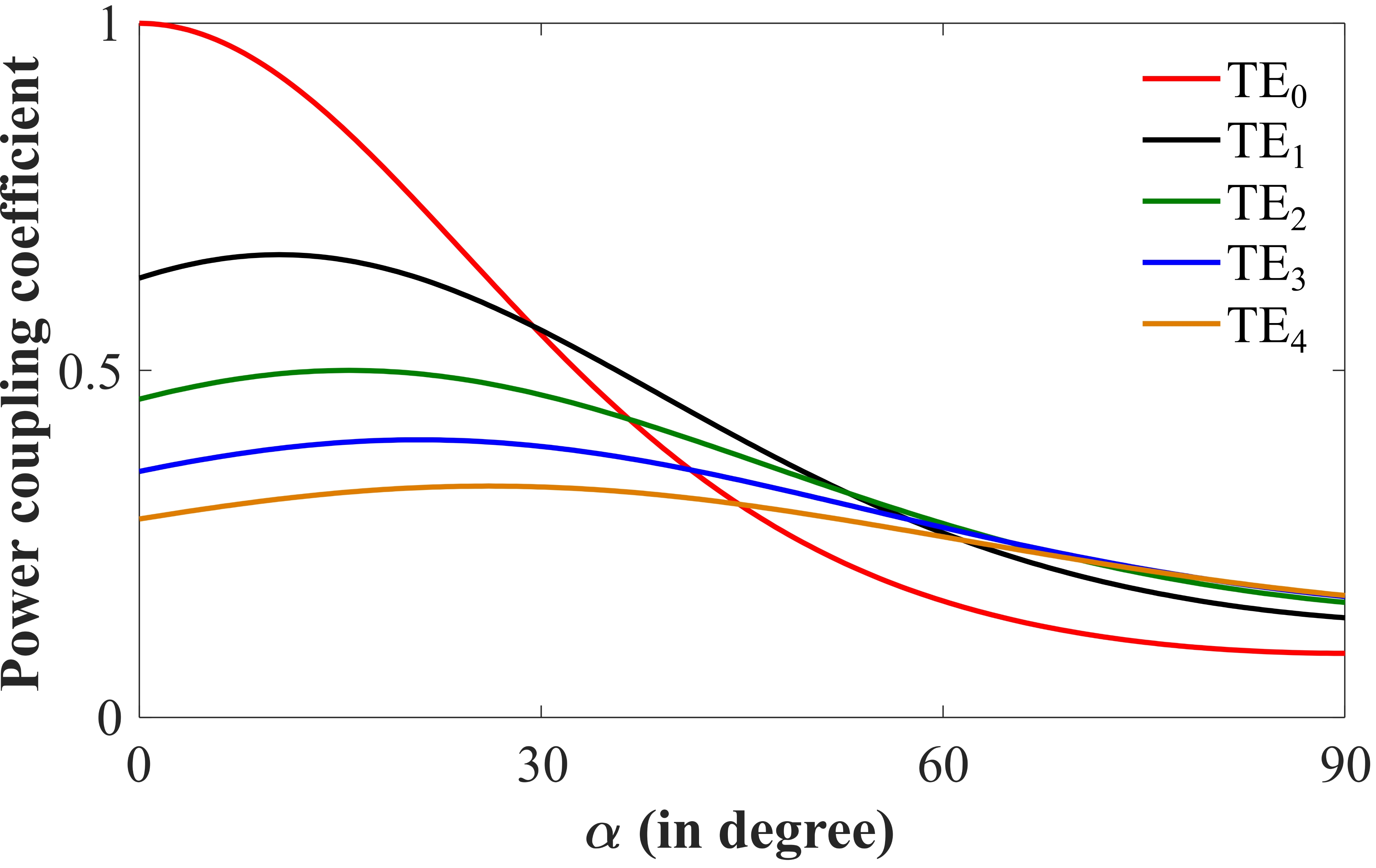}
		\caption{Power coupling coefficients for different TE$_m$ modes as a function of angular misalignment where the input field is TEM$_{00}$.}
		\label{fig4}
	\end{figure}
	
	The power coupling coefficients with respect to $\alpha$ corresponding to different modes  have been calculated  using overlap integral between the launched mode and excited mode, taking into account the diffraction effect of the light beam while propagating in free space due to the inclination between the fiber output end and waveguide axis. In Fig. \ref{fig4}, we have plotted the power coupling coefficients of different modes with respect to $\alpha$ where we observe that on increasing the value of $\alpha$, the power coupling of higher modes reaches a maximum value successively one after another for different values of angle $\alpha$. 
	
    To achieve the targeted fringe spacing, further optimization of $L$ (length of the waveguide) has been taken into account. Hence, a systematic study of light-wave dynamics through the waveguide has been performed. Light wave suffers a small but significant free space propagation before entering into the waveguide due to the presence of finite gap while being coupled into the waveguide through the coupling fiber. The associated electric field of the wave just before entering the waveguide can be expressed as,  
	\begin{equation}
	E_{0}(x,y)=E_{0}\,\exp\left[-ik(x\cos\alpha+y\sin\alpha)\right].
	\end{equation}
	After entering the waveguide, during propagation, the light wave must satisfy the condition, $kd_{co}\sin(\alpha_m)=m\pi$ for quantization of transverse component wave vector ($k_y$). Here, $\alpha_m$ is the angle between $k$- and $x$-axis for $m^{th}$ mode. Whereas, after entering into the waveguide, the wave distribution immediately takes the following form given by \cite{ovchinnikov00}
	\begin{multline}
	E(x,y)=\sum_{m=1}^{\infty}c_m\sin\left(\frac{m\pi}{d_{co}}y\right)\\ \times\exp\left[-x\left\{ik\sqrt{1-\left(\frac{m\pi}{kd_{co}}\right)^2}+\gamma_m\right\}\right].
	\label{Exy}
	\end{multline}
	Here, $c_m$ takes into account the effect of initial distribution of the fields on the $m^{th}$ mode, given that
	\begin{equation}
	c_m=\frac{2}{d_{co}}\displaystyle\int_{0}^{d_{co}}\sin\left(\frac{m\pi}{d_{co}}y\right)E_0(x<0,y)\,dy.
	\end{equation}
	Accordingly, we get transverse momentum of the field distribution inside the waveguide using Fourier transform, which has been expressed as,
	\begin{equation}
		E(k_y)=\frac{1}{d_{co}}\int_{0}^{d_{co}}\exp\left(-ik_yy\right)E(x=L,y)\,dy.
		\label{eky1}
	\end{equation}	
	Using Eq. \ref{Exy}, with the condition $x=L$, Eq. \ref{eky1} can be further elaborated as,
	\begin{subequations}
		\begin{align}
		E(k_y)&=\frac{2E_0}{d_{co}^2}\exp\left(-ixk\cos\alpha,x<0\right)\nonumber\\
		&\times\sum_{m=1}^{\infty}\left[\frac{\left(m\pi/d_{co}\right)^2}{k^2\sin^2\alpha+\left(m\pi/d_{co}\right)^2}\right]\nonumber\\
		&\times\exp\left[-L\left\{ik\sqrt{1-\left(\frac{m\pi}{kd_{co}}\right)^2}+\gamma_m\right\}\right]\nonumber\\
		&\times\left[(-1)^{m+1}\exp\left
		(-id_{co}k\sin\alpha\right)+1\right]^2.\tag{9}
	\end{align}
	\label{eky2}
	\end{subequations}
	It is evident from Eq. \ref{eky2} that, any change in $L$ keeping $d_{co}$ unaltered, will essentially make the momentum distribution at the output to oscillate between $\alpha$ and $-\alpha$. For our proposed geometry we have ${\lambda/d_{co}\ll1}$ and hence ${\sin(\alpha)\ll1}$. To further analyse the wave dynamics, we choose to define dimensionless normalised length ${r=L/S}$, where $S$ is the self imaging period \cite{ovchinnikov01} given by,
	\begin{equation}
	S=\frac{8d^2_{co}}{\lambda}.
	\label{self}
	\end{equation}
	Moreover, to identify the length scales of self imaging period and length required for transverse momentum beating, we define a new propagation parameter ${p=S/S_b}$. Here, ${S_b}$ refers to the period of light momentum beating that is given by (as ${\lambda/d_{co}\ll1}$)	
	\begin{equation}
	S_b=\frac{2d_{co}}{\sin(\alpha)}.
	\label{beat}
	\end{equation}
	For our design, $p$ turns out to be 6.	Now, phenomena like full, half or fractional revivals of the initial wave at the output are a direct consequence of modulation in light intensity inside the waveguide due to transverse momentum beating \cite{ovchinnikov01}. For fabrication feasibility  of the proposed sensor, instead of varying the length of the waveguide, we customise the length $L$ at any point where full or half revivals occur. In this direction for the optimized structure, we opt to choose  the length required for full revival where intensity distribution is essentially the self image of the initial wave. Accordingly, we set $d_{co}$ for a given $\lambda$ in such a way that $p$ turns out to be an integer \cite{ovchinnikov01}. Hence we set the value of $r$ to be an integer while optimizing the length of the device for full revival. This essentially leads to an optimized device length $L=111.5\,mm$.

	
	Now, loss ($\ell$) suffered by different modes can be obtained from,
		\begin{equation}
	      \ell=4.34\left(\frac{k}{2}\right)gL,
		\label{loss}
	\end{equation}	
	where, $k$ and $g$ represent the wave vector and full width half maxima (FWHM) of the field confinement variation with $n_{eff}$, respectively.
	It is observed from Fig. \ref{fig3}(a) that the FWHMs of the peaks increase with increasing the order of mode; and as a consequence, we encounter comparably higher individual loss. Fig. \ref{fig4} shows that for a fixed value of input angle ($\alpha_{m}$), $m^{th}$ mode assumes a maximum amplitude at the input, but for the same incidence angle, higher order modes have smaller power coupled into the guide along with a higher value of loss making their impact negligible. Here comes the problem of relative loss difference of the two interacting modes, which is the direct consequence of field confinement ratio of the interacting modes. As higher order modes engage greater values of FWHM resulting in higher order loss, confinement factor of the mode with low value of FWHM becomes a driving parameter for proper interaction between the selected modes.  For $111.5\,mm$ length there exists a field confinement ratio of $\sim 7.5:1$ between 0$^{th}$ and 2$^{nd}$ mode.

	Following Eq. \ref{dco}, interaction between chosen modes with higher difference in their order will give better sensitivity. Since, the selection of higher order modes is limited by higher loss suffered by that mode, we judiciously choose $0^{th}$ and $2^{nd}$ mode for which a suitable intensity contrast is obtained between them for the optimized length. To excite TE$_2$ inside the guide, we keep input angle ($\alpha=\alpha_2$) approximately at 15.65 degree and in accordance with Eq. \ref{dco}, photonic crystal of length $111.5\,mm$ with core width 7.25 $\mu m$  is producing 1.29 $nm$ spatial resolution, when coupled with a single mode polarization maintaining fiber with 5 $\mu m$ mode diameter, operating at a wavelength of 1.32 $\mu m$.
	That is, a change as small as 1.29 $nm$ in the guide width ($d_{co}$) is making the output beam to swing back to is initial position, enabling the sensing action.

	Thus, by virtue of our proposed interferometric device we obtain a 1.29 $nm$ spatial resolution (or, $\lambda/1024$ fringe spacing). Focusing on the interaction between the lower with a higher order mode, keeping in mind that the relative loss should be minimum along with a maximum difference in between the selected modes i.e., maximizing $(v-u)$, we have considered the interaction between 0$^{th}$ and 2$^{nd}$ order modes. Moreover, better spatial resolution or sensitivity can be obtained by systematically customizing the refractive indices and periodicity of the 1D photonic crystal to realize sharp confinement for the lower as well  the selected higher order mode. Due to sharp confinement, the intensity contrast among the modes will be better as a consequence of low loss ($\ell$) of the respective modes.
	We have addressed the fabrication tolerance of the optimized device geometry. 
	In this direction an estimated error of $\pm{0.1\,\mu m} $ in alignment of the core width ($d_{co}$) along with $\pm{0.1\,\mu m}$ fabrication tolerance in device length is expected to affect the sensitivity by $4\%$.
	
	Further improvement in sensitivity can be systematically achieved by analyzing the trade-off between confinement factor of the chosen set of modes and the difference in the order of chosen modes. Thus following Eq. \ref{dco}, interaction between the lowest ($u$) and highest ($v$) possible mode inside the guide yields maximum sensitivity for a given structural parameters, but the trade-off governs the intensity contrast between the selected modes. For $111.5\,mm$ length there exits a field confinement ratio of $\sim 4$, $\sim 7.5$, $\sim 14.7$ in between 0$^{th}$ and 1$^{st}$, 0$^{th}$ and 2$^{nd}$, and 0$^{th}$ and 3$^{rd}$ modes, respectively. Which indicates the existence of depreciation in intensity contrast between the selected pair of modes. It is clear from the power coupling coefficient plot (Fig. \ref{fig4}) that, for large values of input angle ($\alpha$), starting amplitude of all modes at the input decreases significantly. This phenomenon introduces a trade-off for choosing higher order modes that already suffer higher value of loss. Hence, to excite the same, we need to set $\alpha$ at a higher value. So, we must choose the higher order mode ($v$), in such a way that there exists a significant value for power coupling for the lower order mode ($u$) at that input angle corresponding to mode $v$ (see Fig. \ref{fig4}).


    In summary, we have reported a new design of an interferometric sensor, where a slight variation in the guide width using the actuator can precisely control the output beam to oscillate resulting in a sensitivity as high as $\sim \lambda/1024$ for potential sensing/switching action. Here, the optimized multimode waveguide has been chosen to realize such an ultra-high resolution sensing/switching element when implemented using the photonic crystals. This specific design has enabled high confinement factors for the fields facilitating the waveguide to accommodate higher order modes. An optical fiber supporting a TEM$_{00}$ having mode diameter in accordance with the core width of the guide has been used to excite different modes, interaction among which can produce MMI that in turn makes the output beam to oscillate. Further improvement in sensitivity can be achieved deliberately by further optimization of the photonic cladding to support the modes with a higher difference to commensurate with appropriate relative loss. This structure can be used to make a more compact and robust design of such ultra sensitive interferometer based waveguide sensor wherein the sensitivity of the apparatus would depend on the design of the waveguide within which the phenomenon of MMI is carried out.
    Proposed interferometric sensing scheme offers a good scope to be exploited in photonic sensing/ switching and related precision measurement systems.

\vspace{0.5cm}
\section*{Acknowledgments} 
 SS acknowledges the financial support from Dept. of Physics, Indian Institute of Technology Jodhpur. AL and SG acknowledge the financial support from the Science and Engineering research Board (SERB) under Early Career Research Grant [ECR/2017/000491], Ministry of Science and Technology, Govt. of India.

	\bibliography{ss_ref}
	
\end{document}